\documentclass[%
 reprint,
 amsmath,amssymb,
 aps,
 pra,
 floatfix,
 nofootinbib,
 longbibliography
]{revtex4-1}

\newcommand{\ket}[1]{|#1\rangle}
\newcommand{\bra}[1]{\langle#1|}

\newcommand{\ketbra}[2]{\ket{#1}\bra{#2}}

\usepackage{graphicx}
\usepackage[urlcolor=blue, hyperindex, colorlinks, bookmarks=true]{hyperref} 
\usepackage[capitalise]{cleveref} 
\usepackage[utf8]{inputenc}
\usepackage{microtype} 
\usepackage{bbm}

\newcommand{\wbf}{\boldsymbol{\omega}}
\DeclareMathOperator{\erf}{erf}
\DeclareMathOperator{\SWAP}{SWAP}
\DeclareMathOperator{\iSWAP}{iSWAP}
\DeclareMathOperator{\siSWAP}{\sqrt{iSWAP}}
\DeclareMathOperator{\CZ}{CZ}
\DeclareMathOperator{\CNOT}{CNOT}
\DeclareMathOperator{\CPHASE}{CPHASE}
\DeclareMathOperator{\CCPHASE}{CCPHASE}
\DeclareMathOperator{\XY}{XY}
\DeclareMathOperator{\X}{X}

\DeclareMathOperator{\Z}{Z}
\DeclareMathOperator{\RX}{RX}

\DeclareMathOperator{\RZ}{RZ}

\begin{document}

\title{Implementation of the XY interaction family with calibration of a single pulse}

\author{Deanna M. Abrams}\email[Corresponding author: ]{deanna@rigetti.com}
\author{Nicolas Didier}
\author{Blake R. Johnson}
\altaffiliation[Current address: ]{IBM T J Watson Research Center, 1101 Kitchawan Rd, Yorktown Heights, NY 10598}
\email{blake.johnson@ibm.com}
\author{Marcus P. da Silva}
\altaffiliation[Current address: ]{Quantum Systems Group, Microsoft, One Microsoft Way, Redmond, WA 98052}
\email{marcus.silva@microsoft.com}
\author{Colm A. Ryan}
\affiliation{Rigetti Computing, 2919 Seventh St, Berkeley CA 94701}

\date{\today}

\begin{abstract}
Near-term applications of quantum information processors will rely on optimized circuit implementations to minimize gate depth and therefore mitigate the impact of gate errors in noisy intermediate-scale quantum (NISQ) computers. More expressive gate sets can significantly reduce the gate depth of generic circuits. Similarly, structured algorithms can benefit from a gate set that more directly matches the symmetries of the problem. The $\XY$ interaction generates a family of gates that provides expressiveness well tailored to quantum chemistry as well as to combinatorial optimization problems, while also offering reductions in circuit depth for more generic circuits. Here we implement the full family of $\XY$ entangling gates in a transmon-based superconducting qubit architecture. We use a composite pulse scheme that requires calibration of only a single gate pulse and maintains constant gate time for all members of the family. This allows us to maintain a high fidelity implementation of the gate across all entangling angles. The average fidelity of gates sampled from this family ranges from $95.67 \pm 0.60\%$ to $99.01 \pm 0.15\%$, with a median fidelity of $97.35 \pm 0.17\%$, which approaches the coherence-limited gate fidelity of the qubit pair. We furthermore demonstrate the utility of $\XY$ in a quantum approximation optimization algorithm in enabling circuit depth reductions as compared to the $\CZ$ only case.
\end{abstract}
\maketitle

\section{Introduction}

Noisy intermediate-scale quantum (NISQ) devices have the potential to solve useful problems more efficiently than current classical computing approaches~\cite{Preskill2018}. It is expected that the depth of implementable circuits is an important determinant of a NISQ device's computational power~\cite{Bravyi2019, McClean2018}. For these pre-fault-tolerant devices, the depth to which one can compute is limited by the accumulation of errors from imperfect individual gates; and indeed, most early demonstrations of hybrid quantum-classical algorithms, such as VQE or QAOA~\cite{Otterbach:2017,McCaskey:2019}, have been limited by gate error rates.  Furthermore, in systems with restricted connectivity between qubits, many $\SWAP$s may be required to interact distant qubits, therefore imposing stronger limits on the fidelity of algorithms implemented in these NISQ systems. There are two paths to tackling this scaling problem: one is to drive down error rates in order to increase possible circuit depth; the other is to decrease circuit depth by leveraging a more expressive gate set. 

Typically, arbitrary single qubit rotations have very low error rates and are almost ``free". However, two-qubit entangling gates generally have much larger error rates~\cite{Hong:2019, KraglundAnderson:2019, ibmq:2018, Barends:2014, PhysRevA.93.060302, Barends:2019}, and thus their use should be minimized. When a larger variety of entangling gates is available, it often becomes possible to compile a given circuit containing non-native unitary operations down to fewer native two-qubit gates~\cite{Peterson:2019}. As an example, consider compiling an arbitrary two-qubit gate with some fixed set of entangling gates. Any two-qubit gate can be expressed with at most 3 controlled-$\Z$s ($\CZ$s) or 3 $\iSWAP$s (in addition to single qubit rotations). So while having one of $\iSWAP$ or $\CZ$ does not offer any compilation advantages in the worst case, the average circuit depth of generic circuits will benefit greatly from the availability of both --- e.g., an $\iSWAP$ requires two $\CZ$s and vice-versa. Furthermore, the generic SWAP gate can be implemented using the combination of one $\CZ$ and one $\iSWAP$ rather than 3 $\CZ$s or 3 $\iSWAP$s~\cite{Schuch2003}. This reduction in SWAP overhead is particularly relevant in architectures with nearest-neighbour interactions, as generic NISQ algorithms will require data movement through SWAPs in these architectures. 

$\iSWAP$ is a member of a larger family of $\XY$ gates. $\XY$ gates can be thought of as a coherent rotation by some angle between the $\ket{01}$ and $\ket{10}$ states. Thus, we can define the $\XY$ unitary as
\begin{align} 
\XY (\beta, \theta) =
\begin{pmatrix}
1 & 0 & 0 & 0 \\
0 & \cos(\frac{\theta}{2}) & i \sin(\frac{\theta}{2}) e^{i\beta} & 0 \\
0  & i \sin(\frac{\theta}{2}) e^{-i\beta} & \cos(\frac{\theta}{2}) & 0 \\
0 & 0 & 0 & 1
\end{pmatrix}, \label{unitary}
\end{align}
where $\theta$ gives the rotation angle and depends on the time and strength of the interaction, and $\beta$ gives a phase to the interaction which may be controllable depending on the implementation. We define $\XY(\theta)=\XY(0,\theta)$, and note that $\XY(\pi)=\iSWAP$, and $\XY(\frac{\pi}{2})=\siSWAP$~\cite{PhysRevLett.87.257905, 2001quant.ph.12013K, 2001quant.ph.12025E, Schuch2003}.

If given access to the full family of variable strength $\XY$ entangling gates, even more impressive reductions in average gate depth of $\sim30\%$ are possible for generic circuits, counting only two-qubit entangling gates~\cite{Peterson:2019}. The advantages of particular entangling gates become even clearer when they are well mapped to the problem structure. For example, excitation-preserving operations, such as the $\XY$ family---including $\iSWAP$---are of particular interest for VQE~\cite{2019PhRvP..11d4092G} because they directly emulate dipole-dipole interactions between orbitals. Similarly, excitation-preserving operations also ensure searches for the solution of some combinatorial problems remain in the feasible space~\cite{Hadfield:2019, Wang:2019}.

In this paper we present an implementation of $\XY (\beta, \theta)$ in a superconducting qubit architecture. We validate high fidelity gates for arbitrary $\theta$ using a randomized benchmarking scheme. We are able to achieve these results by focusing on a gate decomposition strategy that leverages precise control over the phase $\beta$ in order to achieve arbitrary entangling strength while only requiring calibration of a single pulse. In order to attain precise control over $\beta$, we provide several circuit identities that describe how to properly track and update $\beta$. These updates are necessary in systems where single qubit control is performed in abstract oscillating rotating frames~\cite{Knill:2000} with $\Z$ rotations implemented as frame updates. Specifically, because the $\XY$ interaction does not commute with local $\Z$ rotations on single qubits, implementation of an $\XY$ gate requires careful treatment of the interactions of these abstract frames. We then present an experimental demonstration of a NISQ algorithm that is able to take advantage of the decreased circuit depth made possible with $\XY$ gates. Lastly, we describe how our decomposition scheme generalizes to several multi-qubit gates of interest.

\section{Parametric XY Interaction}

\begin{figure}
    \includegraphics[width=\linewidth]{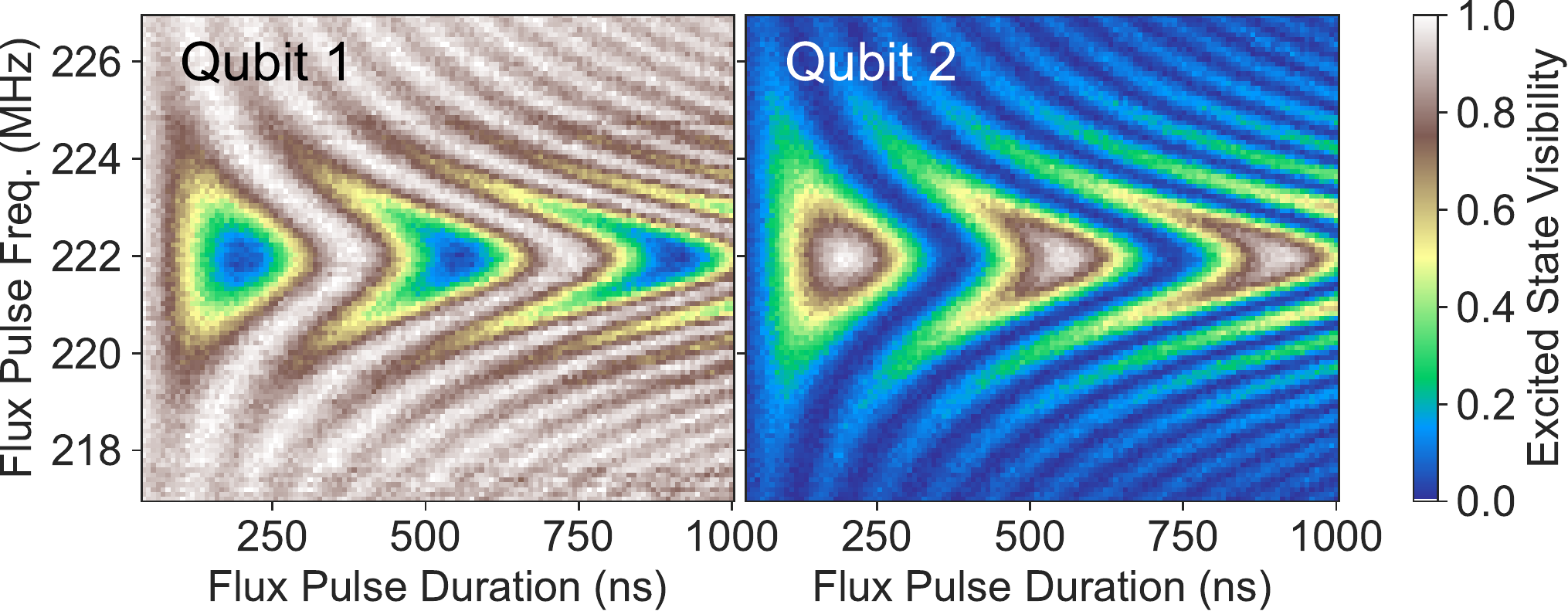}
    \caption{Experimentally measured chevrons associated with the $\XY$ interaction. Plots show population transfer between qubits 1 (fixed frequency) and 2 (tunable frequency) as a function of the duration and frequency of a modulated flux pulse applied to qubit 2 once the state $\ket{10}$ has been prepared. See \cref{sec:experimental_results} for more details on the physical setup.}
    \label{fig:chevrons}
\end{figure}

The $\XY$ interaction corresponds to coherent population exchange between the $\ket{01}$ and $\ket{10}$ states, and thus can be directly generated in systems where a particular form of the anisotropic exchange Hamiltonian is naturally available. A particular system exhibiting the requisite Hamiltonian is that of superconducting qubits with flux tunability. For example, directly bringing two coupled qubits into resonance turns on this interaction \cite{Strauch2003, Majer:2007, Dewes:2012, Barends:2019}. It is also possible to effectively bring coupled qubits into resonance using a modulation sideband from frequency modulating a flux-tunable qubit \cite{Bertet2005, Niskanen2007, McKay:2016, Roth:2017, Didier:2018, Caldwell2018, Mundada:2018, Naik2018}. In particular, for the setting of a fixed-frequency qubit coupled to a tunable-frequency qubit parked at a DC sweet spot and flux-modulated at frequency $\omega_p$, the effective Hamiltonian in the appropriate interaction picture~\cite{Didier:2018} contains the term 
\begin{align}
\hat{H}_\mathrm{int}=g_{\text{eff}}~e^{i(2\omega_p-\Delta)t}e^{i\beta}\ket{01}\bra{10}+\mathrm{h.c.} \label{hamiltonian}
\end{align}
with the notation $\ket{F T}$ for the state $F$ of the fixed-frequency qubit, and $T$ of the tunable qubit. The frequency difference between the two transmons is $\Delta=\omega_{F_{01}}-\omega_{T_{01}}$, $g_{\text{eff}}$ is the effective coupling between the two qubits during modulation, and the phase of the gate is $\beta\approx2\phi_p$ (derived in \cref{app:phase_of_xy_pulse}), where $\phi_p$ is the phase of the flux pulse used to activate the gate, and is defined with respect to the phases of the individual qubit frames. When $\omega_p$ matches the resonance condition (see \cref{fig:chevrons}) we have the desired Hamiltonian that will drive the unitary for $\XY$ defined in \cref{unitary}, up to a sign convention for $\theta$.

\section{Rotating Frame Considerations}

In our experiment, the qubits' single and two-qubit gates are activated by shaped microwave and radio-frequency control pulses. As with many quantum information processing implementations, we define single qubit operations in a local rotating frame oscillating at the mean qubit frequency\footnote{Ideally, one would set the frame frequency to the qubit frequency averaged over possible state preparations of all spectator qubits.}, and implement $\Z$ rotations as updates to the frame definition affecting the phase of subsequent microwave pulses. We also define an equivalent abstract reference frame for the radio-frequency flux pulse that drives the $\XY$ gate, oscillating at the difference frequency of the two single-qubit frames, which we refer to as the ``two-qubit frame".

The reference phase of any given frame can be arbitrarily set to zero when taken in isolation, because our measurement operation measures along the $\Z$-axis and is therefore agnostic to overall $\Z$ rotations. However, the relative phase between frames can have a measurable effect when the corresponding qubits undergo an interaction that does not commute with $\Z$ rotations, and must be accounted for in those cases. 

Importantly, the $\XY(\beta, \theta)$ gates do not commute with single qubit $\Z$ rotations, and thus the relative phase between the single-qubit frames and the two-qubit frame must be accounted for. The phase $\beta$ in \cref{hamiltonian} is defined in an interaction frame given by the doubly rotating frame at the qubit frequencies, and so picks up a linear time dependence on the difference in single qubit frame frequencies. The flux pulse phase in the lab frame must track this time dependence to ensure the same gate is effected no matter when in the pulse sequence it is played. If there are single qubit $\Z$ rotations applied as frame updates, then the definition of $\beta$ must also be updated. This dependence takes the form of an equivalence between the $\XY$ phase $\beta$ and equal and opposite local single-qubit $\Z$ rotations applied before and after the $\XY$ unitary,
\begin{equation}
\includegraphics[scale=0.75]{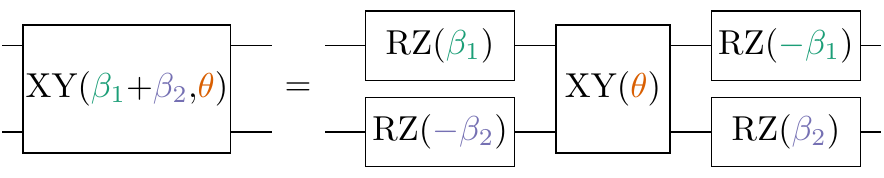} \label{phase_relation}
\end{equation}
where $\RZ$ gates are rotations about $\Z$ on a single qubit, and we use the convention that time flows from left to right (i.e., leftmost gates are applied first). \Cref{phase_relation} may equivalently be written as 
\begin{equation}
\includegraphics[scale=0.75]{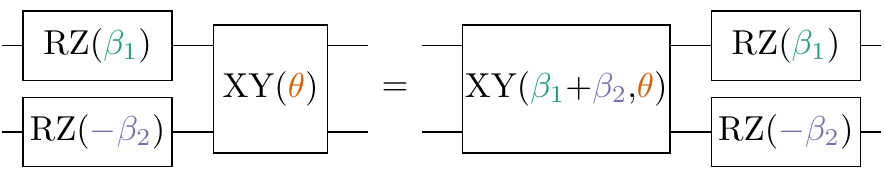}  \label{phase_into_flux_pulse}
\end{equation}
which uses the phase of the flux pulse to leave the single qubit frames untouched after the application of an $\XY$.

\section{\texorpdfstring{$\XY(\beta, \theta)$}{\ifpdfstringunicode{XY(β, θ)}} Decomposition}

The immediately apparent method for tuning up different $\XY(\theta)$ gates would be to vary the interaction time~\cite{2019PhRvP..11d4092G} or strength~\cite{Yan2018} in order to achieve different gate angles. However, this method means that each $\XY$ gate would require a different calibration and/or gate time. In the case of varying gate time, very small rotations may be difficult to achieve given the hardware limitations involved with creating extremely short pulses. Calibrations may be further complicated by nonlinear effects due to hardware constraints when dealing with short pulses, or nonlinear relationships in tunable coupler schemes. 

Instead, we note that the $\XY(\beta,\theta)$ unitary acts non-trivially only in the 2-dimensional subspace spanned by $\ket{01}$ and $\ket{10}$ (the $01/10$ subspace). 
In a two-dimensional subspace, any unitary can be obtained via rotations around two orthogonal axes in the equivalent Bloch sphere, i.e., one can use an Euler decomposition of the unitary. For an arbitrary $\X$ rotation in the Bloch sphere equivalent for a two-dimensional space, one can first absorb the phase $\beta$ in an $\RZ$ sandwich, $\RX(\beta,\theta)=\RZ(-\beta)\RX(\theta)\RZ(\beta)$\footnote{We shirk the common convention for matrix multiplication (rightmost gates are applied to the state first) in order to maintain the same convention as our circuit diagrams (leftmost gates and matrices are applied to the state first).}; and then one can decompose the variable rotation angle $\theta$ as two $\X$ rotations of $\frac{\pi}{2}$ with different phases and a frame update,  $\RX(\theta)=\RX(-\frac{\pi}{2},\frac{\pi}{2})\RX(\frac{\pi}{2}-\theta,\frac{\pi}{2})\RZ(\theta)$.
We then note that an $\RX(\beta,\frac{\pi}{2})$ rotation in the $01/10$ subspace corresponds to $\XY(\beta,\frac{\pi}{2})$ in the full four-dimensional space\footnote{Thanks to \cref{phase_relation}, $\XY(\beta,\frac{\pi}{2})$ is locally equivalent to a $\siSWAP$.}, while an $\RZ(\theta)$ rotation in the $01/10$ subspace corresponds to $\RZ(\frac{\theta}{2})\otimes\RZ(\frac{-\theta}{2})$ in the full four-dimensional space.
This results in
\begin{equation}
\includegraphics[scale=0.75]{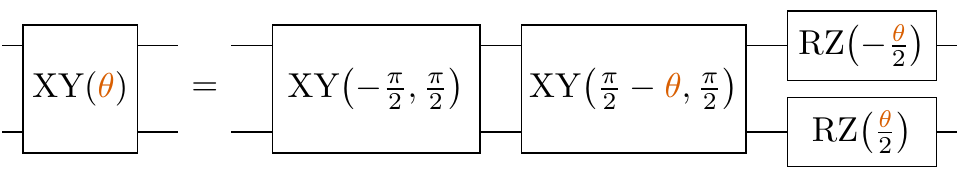}\label{identity}
\end{equation}
which we can use to enact arbitrary $\XY(\theta)$ using only $\siSWAP$ pulses of a controllable phase, and $\Z$ rotations in the single-qubit frames. This provides the advantage of constant gate time as well as a single pulse to calibrate, making it reasonable to fine-tune parameters using coherent error amplification techniques~\cite{Vaughan1972, Kimmel2015}. The only thing changing between different $\XY(\theta)$ gates is the phase of the flux pulse, which allows for a fixed error rate over all $\theta$ in systems where the errors are dominated by a constant decoherence time scale.

It is important to note that in order to make use of ~\cref{identity}, we must be able to apply $\XY(\beta, \theta)$ at a specific $\beta$ in order to perform the intended rotation. That is, we will have to exactly know the phase difference between the two-qubit frame and the single qubit frames. The reason for this becomes clear when thinking of the $\XY(\beta, \theta)$ gates as rotations about a Bloch sphere in the $01/10$ subspace. $\RX(\theta)$ rotations in the single qubit Bloch sphere can be visualized as rotations about an equatorial axis of azimuthal angle 0, by some amount $\theta$. In our pulse decomposition for $\XY$, we think of the $\XY(\beta, \theta)$ gate as an $\X$ rotation in the $01/10$ sphere about an equatorial axis by some amount $\theta$. However, in the case of the $\XY$ gate, the azimuthal angle of the equatorial axis about which we are rotating is set by $\beta$. Because we use the $\XY(\beta,\frac{\pi}{2})$ gate, for which offsets in $\beta$ cannot be corrected by single qubit frame updates applied only after the gate (\cref{phase_relation}), we must be able to set $\beta$ to 0 to complete the analogy between $\RX(\theta)$ and $\XY(\beta, \theta)$, which is what allows us to make use of the Euler decomposition.

\section{Experimental Results}\label{sec:experimental_results}

\begin{figure}
    \includegraphics[width=\linewidth]{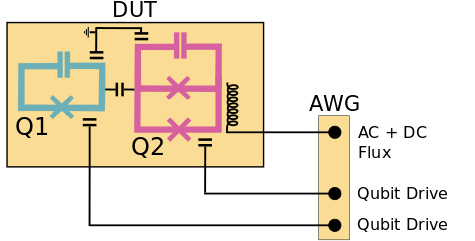}
    \caption{A sixteen qubit chip is placed at the bottom of a dilution refrigerator, with attenuation and filtering similar to that in Ref. \cite{Hong:2019}. In this study, we focus on qubits 1 and 2 (denoted Q1 and Q2), which are a fixed-frequency transmon and a flux tunable-frequency transmon respectively. Qubits 1 and 2 are capacitively coupled to each other, and also capacitively coupled to the ground plane of the chip. The custom AWG has a dedicated channel for each qubit that sends microwave pulses in order to enact single qubit gates. The AWG also has a dedicated channel that combines DC and AC current sources in order to tune Q2, and to enact parametric entangling gates between Q1 and Q2. In order to operate the $\XY$ gate, the AWG maintains phase coherence across all channels shot-to-shot.}
    \label{fig:schematic}
\end{figure}

The experiments in this paper were conducted on two qubits within a larger 16-qubit lattice of superconducting qubits. The two qubit system consists of two capacitively coupled transmons, one with a fixed frequency, and the other with a tunable frequency, referred to as qubits 1 and 2 respectively. The qubits each have their own coplanar waveguide readout resonator coupled to a common readout line. Each qubit is individually controlled by a capacitively coupled RF drive line. The tunable qubit is tuned by an inductively coupled flux line, which is also used to activate the $\XY$ interaction. A block diagram of the experimental setup is shown in \cref{fig:schematic}, and relevant device parameters are listed in \cref{tab:device_table}.

\begin{table*}
\begin{ruledtabular}
\begin{tabular}{r | c c c c c c c c}
& Qubit $f_{01}$ (GHz) & Tunability (GHz) & $T_1$ ($\mu$s) & $T_2^*$ ($\mu$s) & $\tilde{T_1}$ ($\mu$s) & $\tilde{T_2^*}$ ($\mu$s) & Readout Fidelity (\%) & 1Q Gate Fidelity (\%) \\
\hline
Q1 & 3.821 &  --  & 27 $\pm$ 2 & 11 $\pm$ 1 & 24 $\pm$ 2 & 13 $\pm$ 1 & 93.8 $\pm$ 0.1 & 99.83 $\pm$ 0.01 \\
Q2 & 4.759 & .767 &  7 $\pm$ 4 & 8  $\pm$ 1 & 26 $\pm$ 4 & 14 $\pm$ 1 & 96.3 $\pm$ 0.1 & 99.84 $\pm$ 0.02 \\
\end{tabular}
\end{ruledtabular}
\caption{Relevant parameters for the device under test. We list here the qubit transition frequencies ($f_{01}$), the qubit coherence times ($T_1$, $T_2^*$), the qubit coherence times when a flux pulse is applied to Q2 ($\tilde{T_1}$, $\tilde{T_2^*}$), the non-simultaneous readout fidelity, and the simultaneous single qubit gate fidelity as measured by randomized benchmarking. For Q2, which is a frequency-tunable transmon, the qubit frequency was measured at 0 flux, and we additionally list the qubit tunability. 1Q Gate Fidelity is measured using randomized benchmarking of the two qubits in parallel, and the quoted error bars are $1\sigma$ errors on the fit to the decay curve. To extract a readout fidelity, we take the average of the error probabilities for the 0 and 1 states, measured individually. Error bars for coherence measurements as well as the readout fidelity reflect the standard error of the mean over several individual measurements.}
\label{tab:device_table}
\end{table*}

\begin{figure}
    \includegraphics[width=\linewidth]{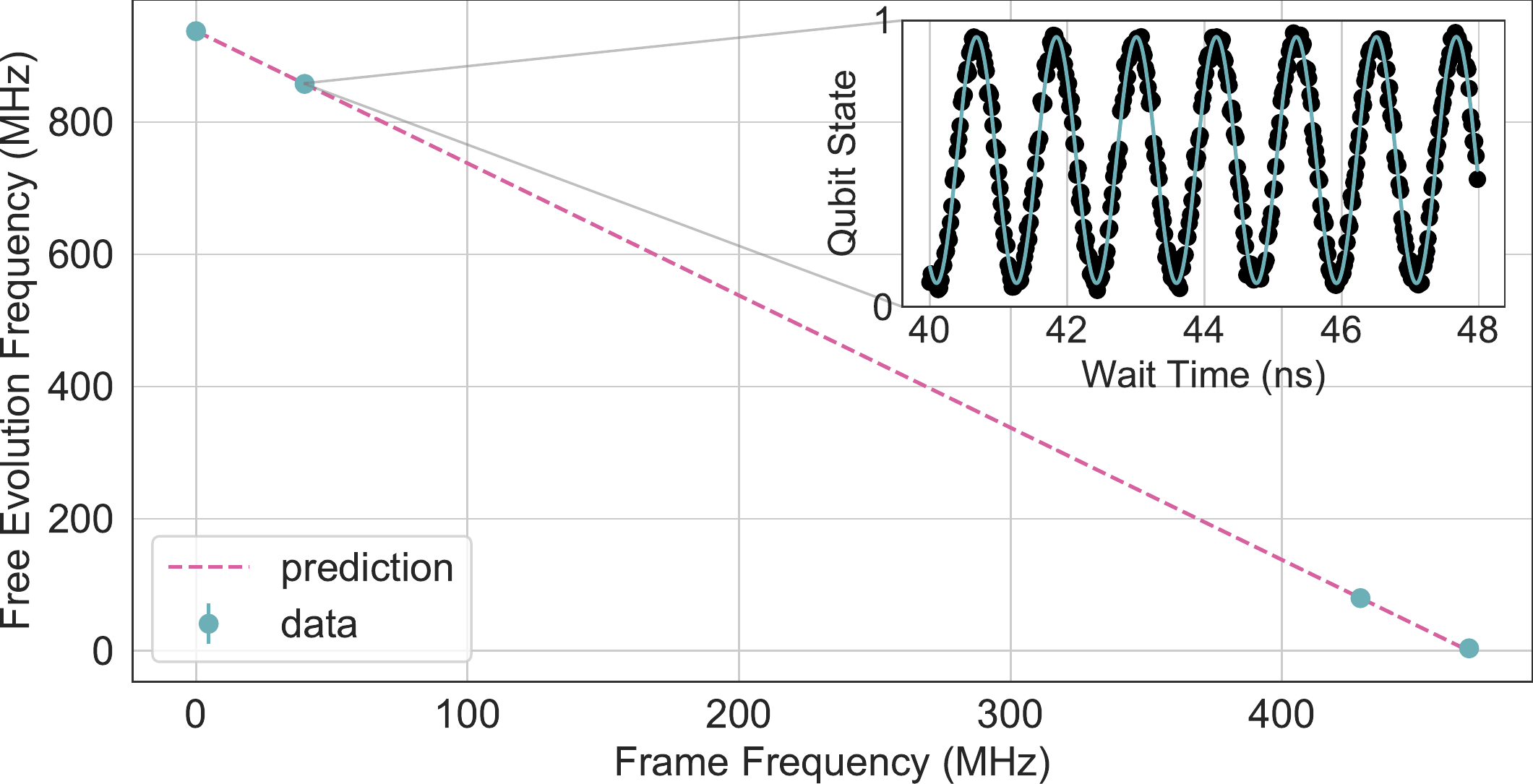}
    \caption{For each data point, the frame of the flux pulse is set to a different frequency, $f_f$, (note that the frequency of the flux pulse itself does not change). In order to measure the free evolution frequency of the equal superposition (with arbitrary phases) of $\ket{01}$ and  $\ket{10}$, a Ramsey-like experiment is performed wherein Q2 is excited to the $\ket{1}$ state; then two $\XY(\beta_0,\pi/2)$ pulses are applied, with a variable wait time between the two flux pulses; then Q2's state is measured. The difference in $f_{01}$ between qubits 1 and 2 is $937.76$\,MHz, and we therefore expect the evolution of the phase of the superposition relative to the two-qubit frame frequency to occur at that frequency. Because $\beta$ goes like twice the phase of the flux pulse (see \cref{app:phase_of_xy_pulse}), we expect that the evolution frequency of the relative phase of the superposition will go like $937.76 - 2 f_f$, and the data agrees very well with this prediction. The inset shows the state of Q2 as a function of the wait time between the two $\XY(\beta_0,\pi/2)$ pulses when the frame frequency is set to $40$\,MHz. We extract the free evolution frequency by fitting a sinusoidal function to Q2's evolution. When the frame frequency is $40$ MHz, we measure the free evolution frequency to be $857.79 \pm 0.27$\,MHz, which agrees with the predicted frequency of $857.76$\,MHz.}
    \label{fig:frame_tracking}
\end{figure}

As discussed in the previous section, the phase of the flux pulse sets the phase of the $\XY$ interaction. However, this phase is defined in an abstract rotating frame whose frequency is set by the frequency of the rotating frame the Hamiltonian is defined in (see \cref{hamiltonian}), which in this case is the difference of the qubit frequencies. We can confirm that the free evolution frequency is what we expect, and that we can track it with an abstract rotating frame, using a Ramsey style experiment in the $01/10$ manifold. We first prepare the $\ket{01}$ states and then do an $\XY(\beta_0,\pi/2)$ (which is equivalent to $\siSWAP$ with unknown flux pulse phase $\beta_0$) to place the state on the equator of a Bloch sphere whose poles are the $\ket{01}$ and $\ket{10}$ states. We let that state evolve and accumulate phase for some time, and then perform a second $\XY(\beta_0+\beta_1,\pi/2)$ pulse to rotate back onto the pole, where the phase $\beta_1$ is a product of the frame frequency and the delay between the two pulses.  
\Cref{fig:frame_tracking} demonstrates that by changing the frame frequency for the flux pulse, we can cancel out the dynamical phase accumulated between the $\ket{01}$ and $\ket{10}$ state, which occurs at the frequency we expect, confirming our model for the abstract rotating frames.

We tune up an $\iSWAP$ on qubits 1 and 2 by applying a flux pulse to qubit 2 and sweeping the frequency, amplitude, and duration of the flux pulse to find the resonance condition for an $\iSWAP$ (see \cref{fig:chevrons}). The pulse time for an $\iSWAP$ is half a period of oscillation (with some offset from non-zero rise and fall times of the pulse). We then additionally calibrate single qubit $\Z$ rotations that are applied after the flux pulse. These allow us to correct both the uncalibrated (but constant) relative phase between the single qubit microwave drives and the modulated flux drive at the chip, as well as the mean shift of the tunable qubit frequency under flux modulation. The reason we can correct for both of these effects with only single qubit Z rotations is that for the special case of $\iSWAP$, similar to a single qubit $\pi$ rotation, \cref{phase_relation} can be simplified to
\begin{equation}
\includegraphics[scale=0.75]{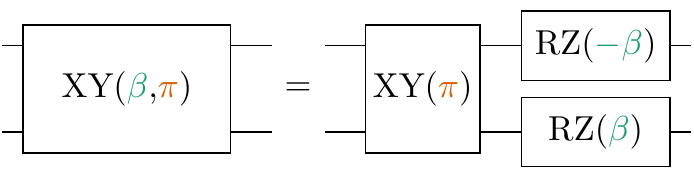} \label{phase_relation_iswap},
\end{equation}
showing that the phase $\beta$ can always be corrected for with post-gate single qubit frame corrections.

\begin{figure}
    \includegraphics[width=\linewidth]{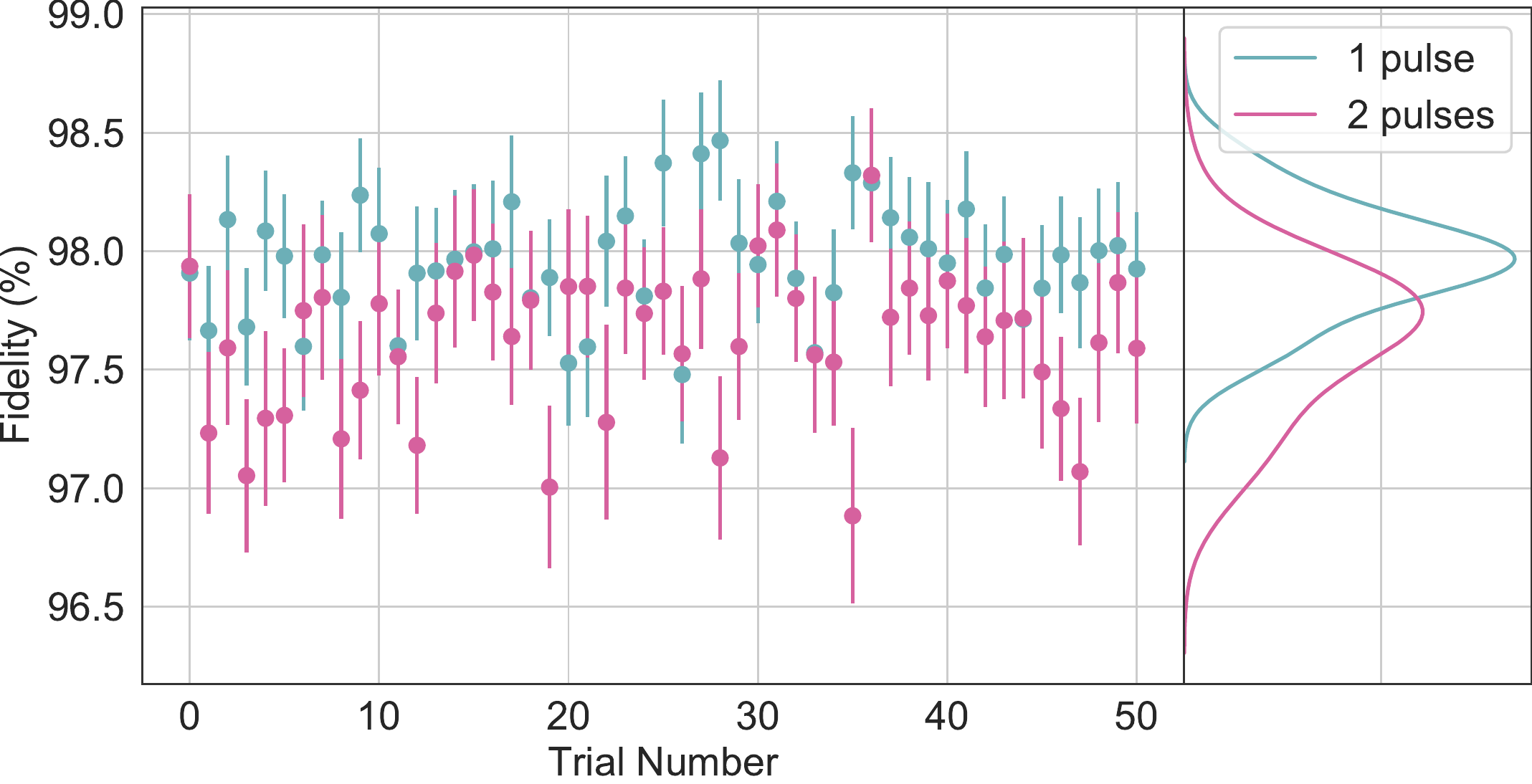}
    \caption{50 iRB measurements were performed, interleaving an $\iSWAP$ gate composed of either a single $\iSWAP$ pulse or two $\siSWAP$ pulses. All of the measurements for a single pulse were performed first, with the two-pulse measurements performed immediately afterwards. Each iRB experiment included 32 trials, at lengths of 2, 4, 8, 16, 32, and 64 Cliffords. Within each individual iRB experiment, the order of the trials was randomized. (left) The iRB fidelity by experiment number. Error bars represent the propagated 1$\sigma$ errors from the fit of the two decay constants. (right) The kernel density estimate of each distribution, showing a slight downward shift for an $\iSWAP$ composed of two pulses compared to a single-pulse $\iSWAP$. The shift in gate fidelity can be explained by coherence limitations given that the total gate time is slightly longer when using two pulses compared to using one.}
    \label{fig:1vs2pulses}
\end{figure}

Once an $\iSWAP$ gate has been calibrated, we benchmark it using interleaved randomized benchmarking (iRB) \cite{Magesan2012}. In order to gather statistics on the gate fidelity and confirm its stability, we perform iRB 50 times (see \cref{app:irb} for a representative iRB experiment, as well as more information on experimental parameters), and measure a median $\iSWAP$ average gate fidelity of $97.98 \pm 0.11 \%$ (see \cref{fig:1vs2pulses}), where errors on the median are derived from the Dvoretzky-Kiefer-Wolfowitz inequality, which gives a 95\% confidence band of the empirical CDF~\cite{DKW,Birnbaum,Massart}. 
In order to test that splitting $\iSWAP$ into two $\siSWAP$ pulses as per \cref{identity} has no adverse effect on gate fidelity, we tune up a $\siSWAP$ pulse by choosing a gate time of a quarter period. When expanding from $\iSWAP$ to $\XY$, the constant phase offset between the single qubit drive and the modulated flux drive can no longer be absorbed in single qubit frame updates, as \cref{phase_relation_iswap} no longer holds and we must fall back to the more general \cref{phase_relation}. Similar to Ref. ~\cite{Naik2018}, we need a suite of phase calibration sequences. 

To calibrate the relative phase between the flux and single qubit drives, we prepare the state $\frac{1}{2}(\ket{00}+\ket{01}+i\ket{10}+i\ket{11})$, perform a single $\XY(2(\phi_p+\phi_0),\pi/2)$, and measure $\langle ZI \rangle - \langle IZ \rangle$, which goes as $\cos(2(\phi_p + \phi_0))$; where $\phi_p$ is the phase of the flux pulse, and $\phi_0$ is the constant phase offset between the single qubit and flux drives. By sweeping $\phi_p$, we can fit the measured expectation value to a cosine, and extract $\phi_0$ in order to correct for it and perform a true $\siSWAP$. 

Additionally, each flux pulse causes the qubits to incur $\RZ$s due to the mean shift of the qubit frequency under flux modulation. The $\RZ$s induced by the first flux pulse we absorb into the phase of the second flux pulse using \cref{phase_into_flux_pulse}. The phase of the second flux pulse relative to the first is calibrated by preparing the $\ket{01}$ state, applying two $\XY(\beta, \frac{\pi}{2})$ flux pulses, sweeping the phase of the second flux pulse, and maximizing the overlap on $\ket{01}$ such that the effective $\beta$s of the two flux pulses have a difference of $\pi$, and the two $\XY(\beta, \frac{\pi}{2})$ coherently cancel to implement an $\XY(0)$ operation.

The sum of the $\RZ$s carried forward from the first flux pulse and those incurred during the second flux pulse we correct for with final single qubit $\RZ$s chosen to zero the phase accumulated on single qubit superposition input states after application of a nominal $\XY(0)$ unitary The single qubit $\RZ$s calibrated in this way are constant with respect to $\theta$ thanks to our constant gate time. 

We then perform 50 runs of iRB using two $\siSWAP$ pulses to enact $\iSWAP$, and compare with the 50 runs of iRB already performed on the single-pulse $\iSWAP$. \Cref{fig:1vs2pulses} plots the $\iSWAP$ gate fidelity for each trial, as well as the kernel density estimate for the single-pulse and two-pulse $\iSWAP$ fidelities. The two-pulse version has a small but measurable decrease in fidelity that we attribute to the increased gate time gained due to finite rise and fall times ($152\,\mathrm{ns} + 152\,\mathrm{ns}=304\,\mathrm{ns}$ for two pulses vs. 240\,ns for a single pulse). Given the measured coherence times for each qubit under modulation conditions similar to those experienced during the $\iSWAP$ gate (see \cref{tab:device_table}), we can calculate the coherence-limited theoretical gate fidelities for each gate time, using similar techniques as those employed in Ref. ~\cite{Didier:2018ss}. For an $\iSWAP$ enacted with one pulse, the coherence limited gate fidelity is $98.15 \pm 0.08\%$, and for an $\iSWAP$ composed of two flux pulses, the coherence limited fidelity is $97.65 \pm 0.1\%$, whereas we observe gate fidelites of $97.98 \pm 0.11\%$ and $97.72 \pm 0.16\%$. The errors on the coherence-limited fidelity calculations are propagated from the uncertainties in coherence times under modulation. We expect that optimal control approaches will enable a gate time for the two-pulse technique equal to the gate time for the single-pulse technique, eliminating this small drop in fidelity.

\begin{figure}
    \includegraphics[width=\linewidth]{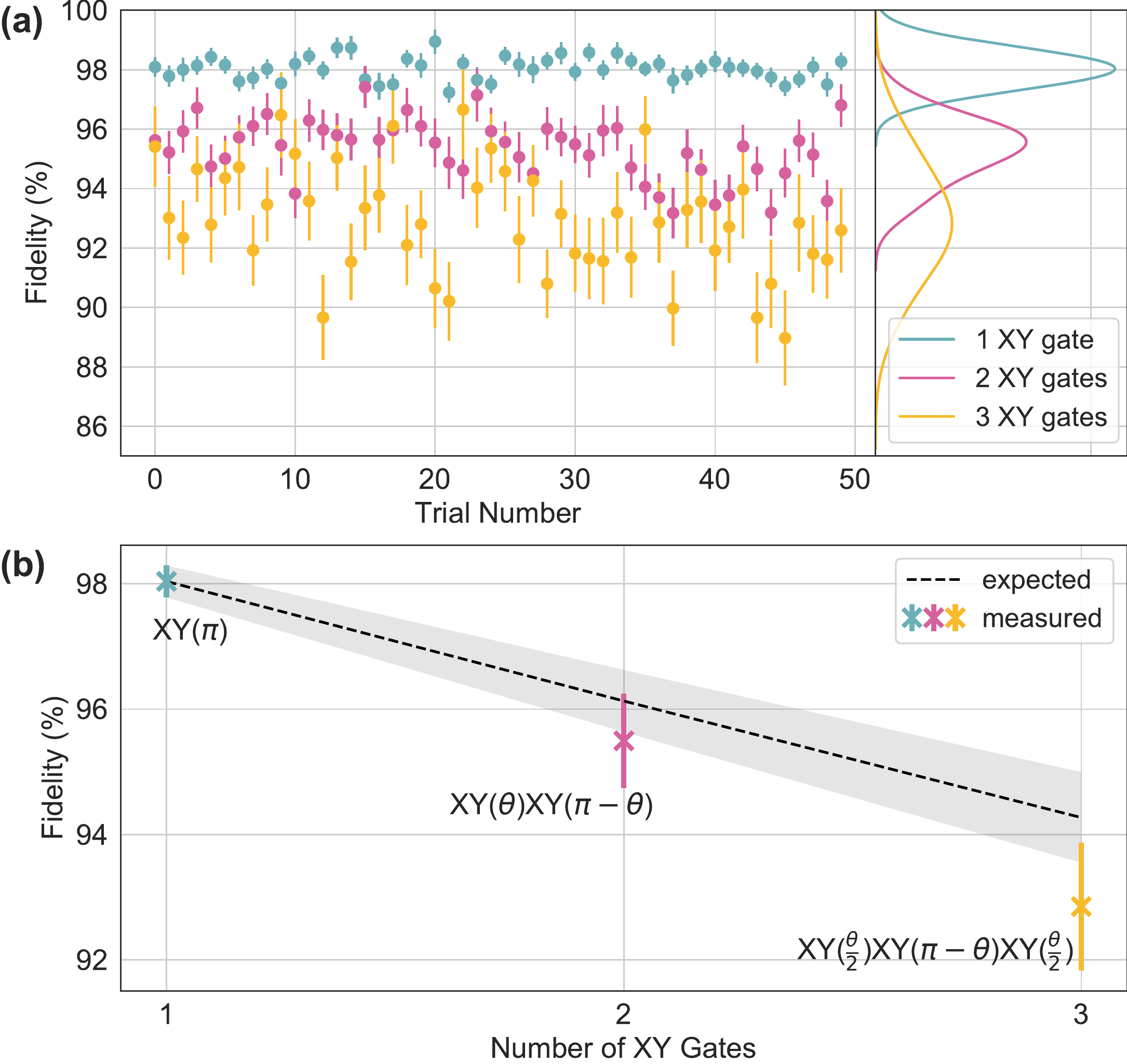}
    \caption{(a) $\iSWAP$ iRB fidelity, composing $\iSWAP$ from 1, 2, and 3 $\XY(\theta)$ gates. Left is the measured fidelity by trial number, where error bars represent the propagated 1$\sigma$ errors from the fit of the two decay constants. Right shows the kernel density estimate of each distribution. We use these measurements to confirm that the median of each distribution scales with the number of $\XY$ pulses used to compose an $\iSWAP$, a fact we will later use to infer the fidelity of $\XY(\theta)$ by performing iRB on $\XY(\theta)$, $\XY(\pi-\theta)$ (details in the text). (b) Median fidelity of each sequence of $\XY$ gates over 50 experiments, with plotted error bars showing the 95\% confidence interval of the empirical cumulative distribution function (ECDF). The median fidelity of a single $\XY$ is $98.04 \pm 0.25\%$. For two $\XY$s, the median fidelity is $95.49 \pm 0.75\%$. For three $\XY$s, the median fidelity is $92.85 \pm 1.02\%$. For a single $\XY$ fidelity of $98.04\%$, we expect two $\XY$s to have a fidelity of $96.13\%$, and three $\XY$s to have a fidelity of $94.27\%$, which is within the spread of what we observe. The gate sequence used for each number of $\XY$s is annotated on the plot, where $\theta$ was chosen randomly each time the $\XY$ sequence was performed.}
    \label{fig:irb_pulse_decomp}
\end{figure}

Once $\siSWAP$ is tuned up, we can smoothly control the relative phase of the second  $\siSWAP$ pulse to achieve any gate in the $\XY(\theta)$ family as per \cref{identity}. We would like to benchmark $\XY(\theta)$ using iRB; however, arbitrary $\XY(\theta)$ gates are not in the Clifford group, which is a requirement of iRB. Instead, we benchmark the gate by building iRB sequences for $\iSWAP$ (a Clifford group operation), and then decomposing $\iSWAP$ into pairs of $\XY(\theta)\XY(\pi-\theta)$, or triples of $\XY(\frac{\theta}{2})$, $\XY(\pi-\theta)$, $\XY(\frac{\theta}{2})$~\footnote{We use this method to avoid the state preparation and measurement errors~\cite{2013PhRvA..87f2119M} endemic to process tomography.}. For iRB, we are directly measuring the average fidelity of the interleaved operation, which in these cases is one, two, or three $\XY(\theta)$ gates. These in turn can be related to the fidelity of a single $\XY(\theta)$, under the assumption that fidelity is independent of $\theta$, i.e. that there are no coherent errors, via the appropriate root of the ratios of the reference decay rate $p$ and interleaved decay rate $p_{il}$ (see \cref{app:irb}). \Cref{fig:irb_pulse_decomp}(a) shows the results of these measurements over 50 rastered runs of each type, and \cref{fig:irb_pulse_decomp}(b) shows the median fidelities for each set of experiments compared to the expected values based on scaling the fidelity of a single $\iSWAP$. The gate fidelity trends with the number of $\XY$ gates used to enact an $\iSWAP$ as we expect, confirming that we can estimate the fidelity of a single $\XY$ gate from an iRB experiment wherein we construct $\iSWAP$ out of $\XY(\theta)\XY(\pi-\theta)$. We note that this approach could hide cancelling coherent errors between the $\XY(\theta)$ and $\XY(\pi-\theta)$ gates; however, as the measured scaling is slightly worse than the expected scaling (see \cref{fig:irb_pulse_decomp}(b)), it is unlikely that we are overestimating the inferred single $\XY$ gate fidelity.

Finally, to verify that we can indeed perform $\XY$ gates for arbitrary $\theta$ with constant fidelity, we perform iRB on $\XY(\theta)\XY(\pi-\theta)$ for 102 randomly chosen $\theta$s, and take the square-root of the decay ratios $\frac{p_{il}}{p}$ (as described in \cref{app:irb}) to extract the fidelity of a single $\XY(\theta)$. Between the angles of $0$ and $2\pi$, we measure a median $\XY(\theta)$ fidelity of $97.36 \pm 0.17\%$, with a range of $95.70 \pm 0.59\%$ to $99.01 \pm 0.15\%$. The lack of obvious correlation between $\theta$ and fidelity confirms constant gate fidelity as a function of $\theta$, proving the utility of our decomposition in unlocking a full family of parametrically controllable entangling gates.

\begin{figure}
    \includegraphics[width=\linewidth]{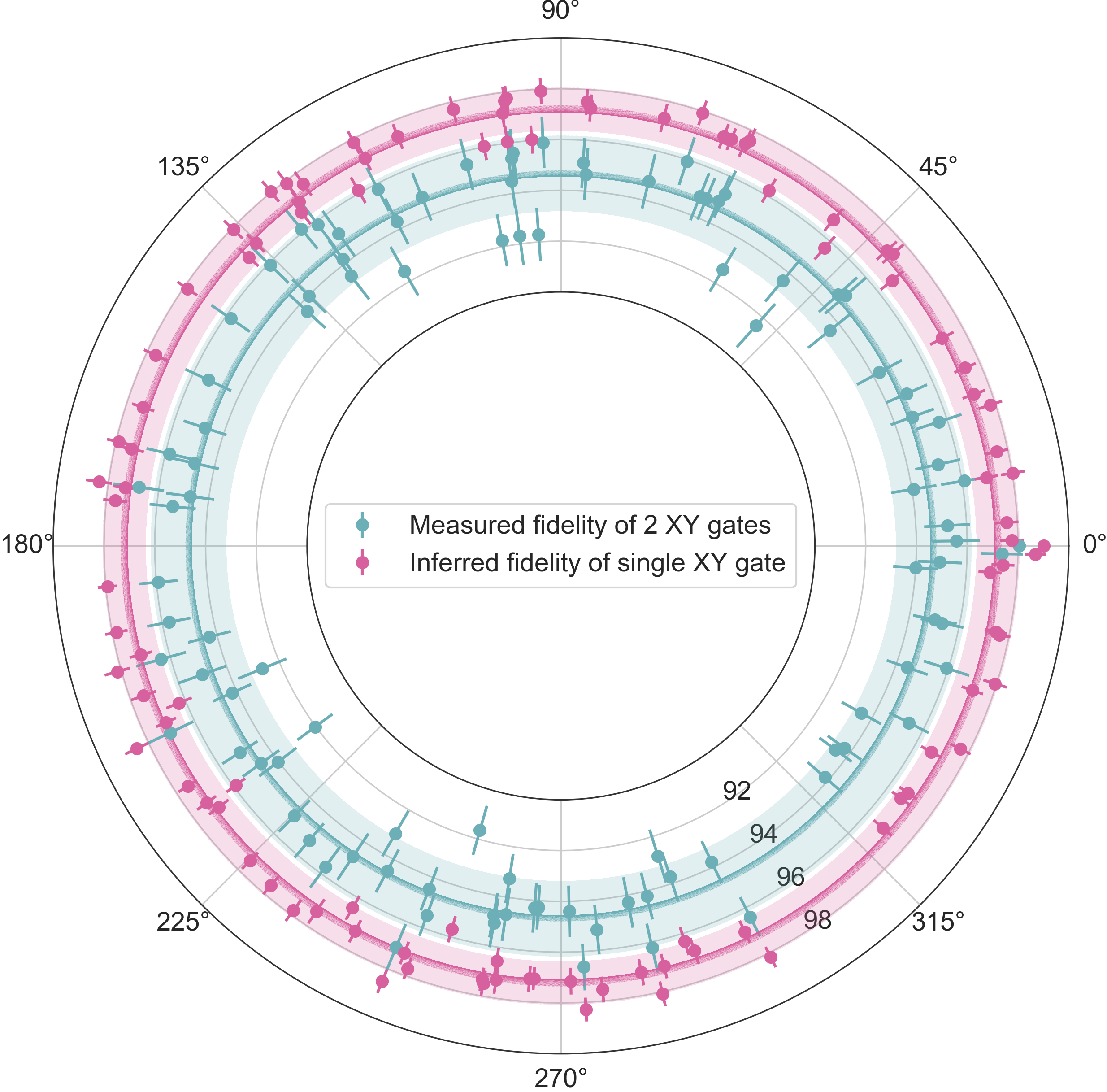}
    \caption{$\XY(\theta)$ $\XY(\pi - \theta)$ fidelity as measured from iRB is plotted on the radial axis as a function of $\theta$ in teal, where $\theta$ is chosen randomly for each of the 102 experiments. In magenta is the inferred fidelity of a single $\XY$ gate at each $\theta$. It is important to note that this inferred fidelity will be due to some average of the $\XY(\theta)$ and $\XY(\pi - \theta)$ errors. Guiding lines are plotted at the median of each distribution, with shaded regions representing the interquartile range.}
    \label{fig:xy_theta}
\end{figure}

\subsection{Reducing Circuit Depth in MaxCut QAOA}

Once both the $\XY$ and $\CZ$ gates are available, it is possible to run algorithms such as QAOA~\cite{farhi2014quantum} with fewer gates. As a demonstration, we calibrate both $\XY$ and $\CZ$ entangling gates between the pairs 0-1, 1-2, and 2-3, which are connected in a line; and additionally take into account phase corrections that must be applied to spectator qubits for each gate due to crosstalk~\cite{Abrams2019}. We then use these qubits to map out a series of weighted MaxCut QAOA landscapes\cite{Otterbach:2017, Zhihui2018, qiang2018, verdon2019learning}. A MaxCut problem seeks to partition a graph of vertices and edges into two sets such that the number of edges in each set is maximized. The encoding of the problem is given by a cost Hamiltonian. QAOA seeks to find the ground state of the cost Hamiltonian by first preparing an equal superposition of all bit strings, and then evolving the system with the cost Hamiltonian followed by a driver Hamiltonian, which are parameterized by some angles $\gamma$ and $\beta$\footnote{Note that this $\beta$ is unrelated to the phase of the $\XY$ gate. Due to its common use in the literature, we suffer the definition collision.} respectively. For the optimal choice of angles, the probability of sampling the bit string which corresponds to the ground state of the cost Hamiltonian increases with the number of alternating applications of the cost and driver Hamiltonians. In weighted MaxCut, each edge of the cost Hamiltonian is assigned a weight chosen at random, making the QAOA landscape dependent on the choice of weights. In our implementation, we merely map out the MaxCut landscape by sweeping $\beta$ and $\gamma$ uniformly. Furthermore, we only apply the cost and driver Hamiltonians a single time, a scenario for which the optimal angles are known.

We experimentally measure two different weighted MaxCut graphs, one graph of four vertices connected in a ring (4 edges), and one graph of four vertices with all-to-all connectivity (6 edges). These MaxCut graph topologies are different from the actual connectivity of the four qubits we use to perform the calculations, which are only connected in a line (0-1-2-3). When mapping the MaxCut graph topology onto the actual device topology, it becomes necessary to introduce SWAPs in the circuit in order to effectively achieve the connectivity required by the MaxCut problem.

To solve weighted MaxCut for a graph with a ring topology, the QAOA circuit can be represented using $\CNOT$s as
\begin{widetext}
\begin{equation}
\includegraphics[scale=0.8]{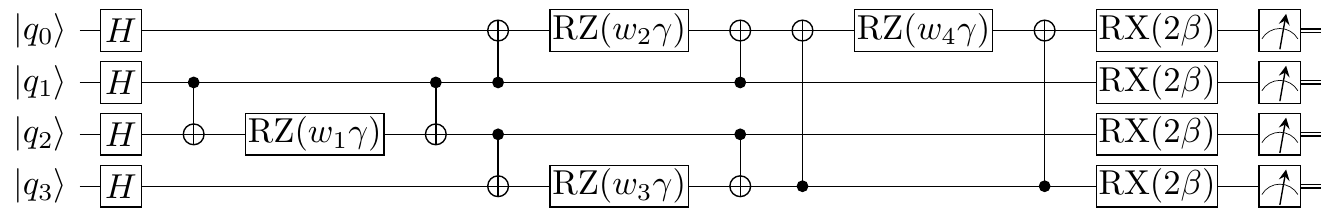}\label{eq:circuit-ring},
\end{equation}
\end{widetext}
where $w_i$ is the weight of the associated edge, and is chosen randomly when generating the circuit. When compiling ~\cref{eq:circuit-ring} to reflect the actual gates available and the actual connectivity of the device, SWAPs will have to be added to account for the limited connectivity of the qubits, specifically for the final phase-gadget~\cite{alex2019phase}. When only $\CZ$ gates are available, the circuit can be compiled to use 10 $\CZ$s. However, when both $\CZ$ and $\XY$ are available, the circuit can be compiled to use only 6 $\CZ$s and 2 $\XY(\pi)$s, giving a gate count reduction of 20\% (the supplementary material contains the actual circuits used to generate the QAOA landscapes~\cite{supplementary}). 

\begin{figure}
    \includegraphics[width=\linewidth]{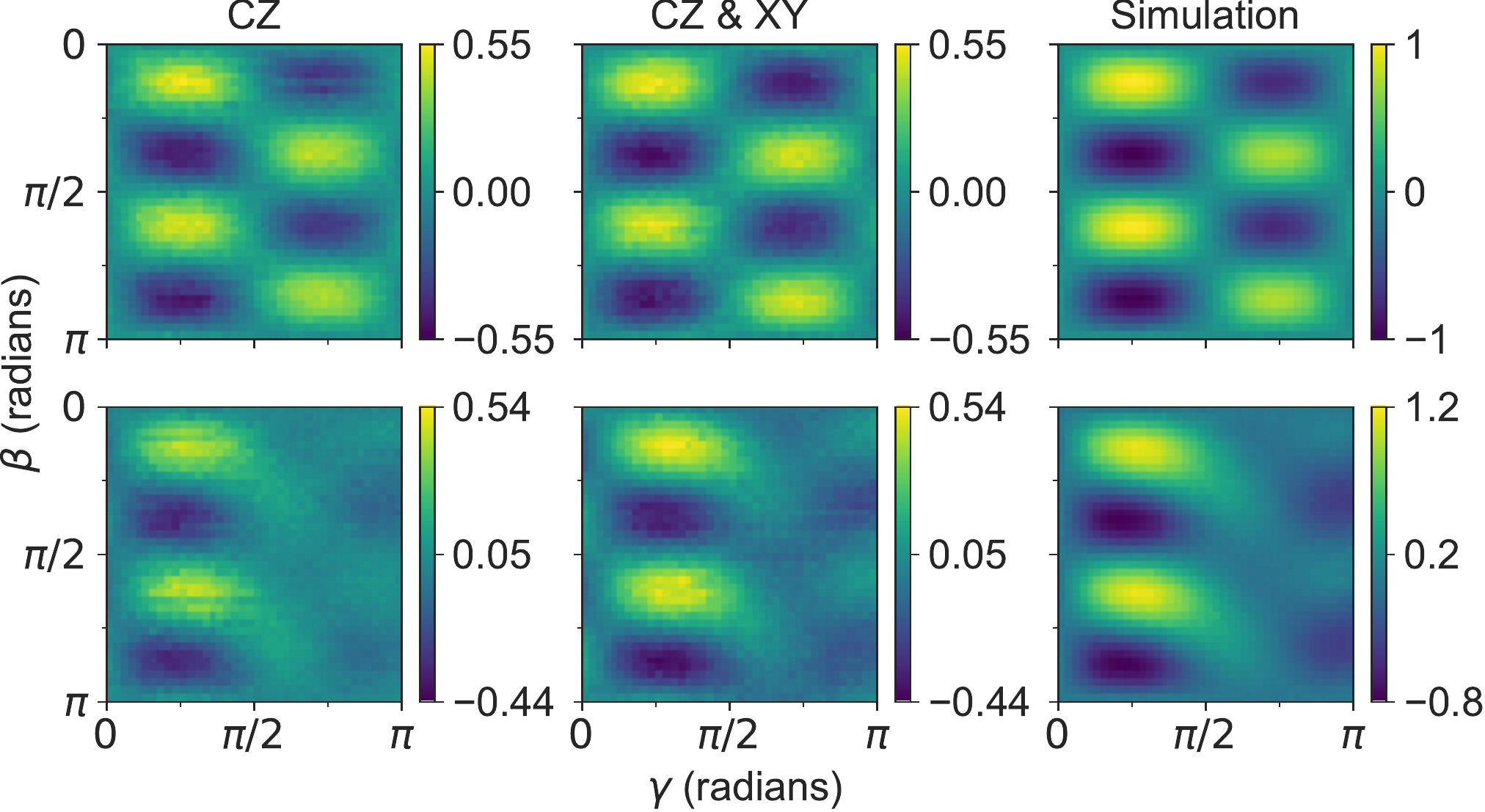}
    \caption{MaxCut QAOA landscapes, showing bitstring cut weights as a function of $\gamma$ and $\beta$. Top row is QAOA on a graph of four edges connected in a ring. Bottom row is QAOA on a graph of four edges with all-to-all connectivity. In both instances the problem is mapped to four qubits with line connectivity. Leftmost column is experimental data using only $\CZ$ gates. Middle column is experimental data using both $\CZ$ and $\XY$ gates. Rightmost column shows simulated results on noiseless qubits. All four experimental landscapes were taken with 5000 shots per angle pair.}
    \label{fig:qaoa}
\end{figure}

For a fully connected MaxCut graph, the QAOA circuit can be represented using $\CNOT$s as 
\begin{widetext}
\begin{equation}
\includegraphics[scale=0.8]{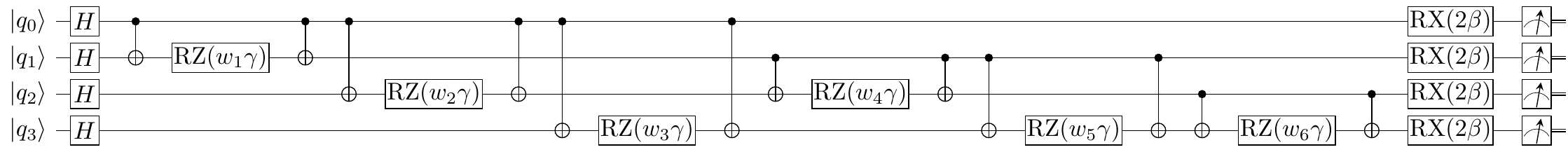}\label{eq:circuit-fully-connected}.
\end{equation}
\end{widetext}
Again, because the connectivity of the actual qubits being used to run the circuit is more restricted than the MaxCut graph, SWAPS will have to be introduced into the circuit in order to actually implement all of the required operations. When only $\CZ$ gates are available, \cref{eq:circuit-fully-connected} can be compiled to use 17 $\CZ$s. However, when both $\CZ$ and $\XY$ are available, the circuit can be compiled to use only 7 $\CZ$s and 5 $\XY(\pi)$s, giving a gate depth reduction of $\sim30\%$ (see the supplementary material for the full circuits used in the experiments~\cite{supplementary}). 
\Cref{fig:qaoa} shows that the QAOA landscapes for circuits compiled with both $\CZ$s and $\XY$s are of similar quality to those compiled using only $\CZ$s in both the fully connected and ring problems. This similar quality, paired with reductions in gate depth, is expected to yield improvements in algorithm performance. 

It is interesting to note that for the special case of combinatorial problems that can be mapped to a cost Hamiltonian consisting of a sum of two body terms (such as MaxCut), the availability of $\iSWAP$ in addition to $\CZ$ allows one to maintain the same two-qubit-gate count and circuit depth on qubits connected in a line as would be possible with the availability of all-to-all connectivity~\cite{babush2018,Kivlichan:2018,crooks2018performance,ogorman2019generalized}.

\section{Extensions to other gate families}

The decomposition method which allows access to the full $\XY(\theta)$ family through tuneup of a single pulse can be easily extended to the $\CPHASE(\theta)$ family. The key observation is that $\CPHASE$ can be implemented equivalently to an $\XY$ interaction, where in the $\CPHASE$ case, the excitation is exchanged between the $\ket{11}$ and $\ket{02}$ states\footnote{Similar results hold for the exchange between $\ket{11}$ and $\ket{20}$.} of a pair of transmons~\cite{Strauch2003,Reed2012,Didier:2018}. This interaction is detailed in the Hamiltonian $H_{\XY,02}(\beta)=e^{i\beta}\ketbra{11}{02}+e^{-i\beta}\ketbra{02}{11}$, which allows for the unitary operation $\XY_{02}(\beta,\theta) = \exp\left[-i\frac{\theta}{2}H_{\XY,02}(\beta)\right]$ (where again we define $\XY_{02}(\theta)=\XY_{02}(0,\theta)$.

A $\CZ$ gate can then be implemented as a $2\pi$ rotation in the $11/02$ subspace. The more general $\CPHASE(\theta)$ gate can be implemented as 
\begin{equation}
\includegraphics[scale=0.75]{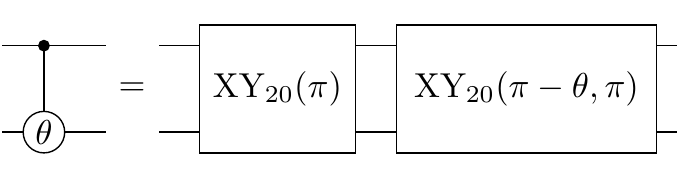}
\end{equation}
where the $\pi-\theta$ phase shift can be realized by changing the relative phase between the two flux pulses used to enact the $\XY_{02}(\pi)$~\cite{Naik2018}. Thus, much like the entire $\XY(\theta)$ family can be implemented by calibrating a single $\XY(\frac{\pi}{2})$ gate and changing phases, the entire $\CPHASE(\theta)$ family can be implemented by calibrating a single $\XY_{02}(\pi)$ gate and changing phases.

The combination of $\XY(\beta,\theta_1)$ and $\CPHASE(\theta_2)$, which can be implemented by tuning only $\XY(\frac{\pi}{2})$ and
$\XY_{02}(\pi)$, leads to the fSim (``fermionic simulation'') family of gates, which are particularly useful in electronic structure calculations~\cite{Kivlichan:2018}.

Interestingly, a Toffoli gate~\cite{Barenco95} can also be implemented via a $\CCPHASE(\pi)$ gate with only 4 applications of $\XY_{02}(\beta,\pi)$ and/or $\XY_{20}(\beta,\pi)$ interactions between three neighbouring transmons (in fact, three transmons arranged in a line)~\cite{Fedorov2012, Mariantoni2011}\footnote{Unlike the work in Ref.~\cite{Fedorov2012}, here we have control over the phase, which allows us to use only $\pi$ pulses, instead of needing $\pi$ and $3\pi$ pulses.}, 
\begin{equation}
\includegraphics[scale=0.7]{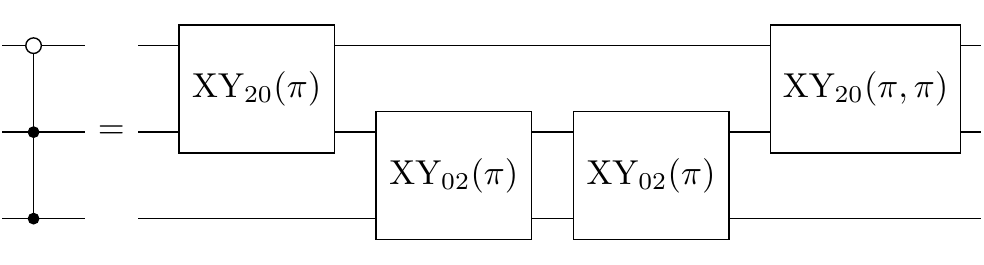}
\end{equation}
More generally, one can implement the whole $\CCPHASE(\theta)$ family~\cite{Barenco95} with only 4 applications of  $\XY_{02}(\beta,\pi)$ and/or $\XY_{20}(\beta,\pi)$, using
\begin{equation}
\includegraphics[scale=0.7]{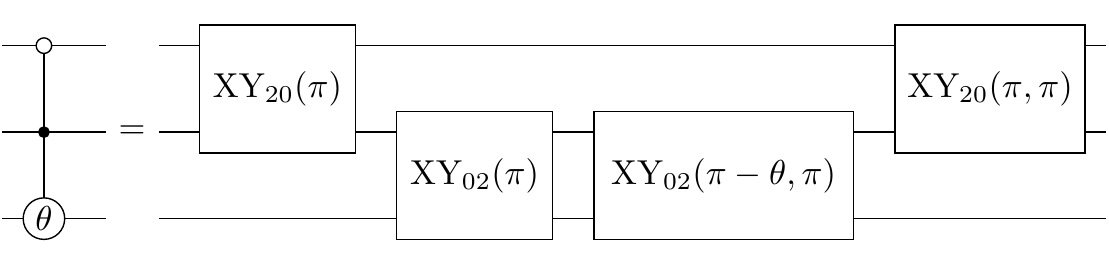},
\end{equation}
which is a small modification of the circuit used in ref.~\cite{Fedorov2012}\footnote{These constructions generalize straightforwardly to more control qubits. The fidelity is expected to degrade due to enhanced relaxation and dephasing rates of the second excited state of the transmon, as well as due to additional error mechanisms (such as state population remaining outside the qubit manifold after an imperfect gate operation). These can be addressed by applying additional pulses to the transmons that may contain higher excitations. That being said, the trade-off to consider is between the greatly reduced circuit depth and the new error mechanisms introduced.}.
Without access to these operations, a Toffoli gate requires as many as 6 $\CZ$s to be implemented on 3 qubits arranged in a triangle~\cite{DiVincenzo1998,Schuch2003}, and as many as 2 $\CZ$s and 6 $\iSWAP$s to be implemented on 3 qubits arranged in a line (a simple extension of results in Ref.~\cite{Schuch2003}). The approach we describe here leads to a 2-3$\times$ reduction in circuit depth for a Toffoli gate, and likely larger improvements for the more general $\CCPHASE$ (as high as 4-6$\times$ based on the construction described in Ref.~\cite{Barenco95}). $\CCPHASE(\theta)$ gates have been shown to be useful in proposed ``quantum supremacy'' demonstrations~\cite{Dalzell2018}, and are also useful in quantum error correction in the absence of measurement \cite{Cory1998, Paz-Silva2010, Schindler2011, Reed2012}.

These results further illustrate that an expressive gate set can have a significant impact on circuit depth, while remaining grounded in an implementation that allows for access to these continuous gate families by tuning up a very small number of actual pulses. The calibration of $\X(\beta,\pi/2)$, $\XY(\beta,\pi/2)$, $\XY_{02}(\beta,\pi)$, and $\XY_{20}(\beta,\pi)$ (where the calibration is independent of $\beta$) gives access to all single qubit rotations, the entire $\XY(\beta,\theta)$ family, and the entire $\CPHASE(\theta)$ family, with only two pulses each. These pulses further allow implementations of the entire $\CCPHASE(\theta)$ family (including $\CCPHASE(\pi)$, which is equivalent to a Toffoli gate) in only 4 pulses.

\section{Conclusion}

In this paper, we have shown high fidelity operation of $\iSWAP$ as well as demonstrated the ability to selectively perform any operation within the $\XY$ gate family by controlling the relative phase between two $\siSWAP$ pulses comprising the entangling gate. This implementation only requires tuning up a single pulse, and maintains constant gate time for all $\theta$, easing pulse shaping requirements on the control system. The median gate fidelity across the sampled $\XY$ gates is $97.35 \pm 0.17\%$, supporting the hypothesis that gate fidelity is insensitive to entangling angle. We have additionally outlined a framework for properly tracking the phases of several abstract rotating frames as they interact through the $\XY$ interaction, and employed hardware able to maintain phase stability across several output channels in order to take advantage of this framework. Furthermore, we demonstrate that MaxCut QAOA gate depth can greatly benefit from the availability of $\XY$ in addition to $\CZ$, which we expect to lead to improvements in its utility as a NISQ algorithm. Finally, we show that our decomposition scheme can be expanded to many different gates, allowing for an expressive gate set through calibration of a small number of pulses.

\begin{acknowledgments}

Work at the Molecular Foundry was supported by the Office of Science, Office of Basic Energy Sciences, of the U.S. Department of Energy under Contract No. DE-AC02- 05CH11231. We thank the Rigetti quantum software team for providing tooling support, the Rigetti fabrication team for manufacturing the device, the Rigetti technical operations team for fridge build out and maintenance, the Rigetti cryogenic hardware team for providing the chip packaging, and the Rigetti control systems and embedded software teams for creating the Rigetti AWG control system. We additionally thank Michael Scheer and Eric Peterson for convincing us that the composite pulse decomposition would be a valid implementation of the $\XY$ gates. We thank Peter Karalekas for his help in setting up the QAOA demo. Lastly, we thank Zhihui Wang, Nicholas C. Rubin, Eleanor G. Rieffel, and Davide Venturelli for valuable conversations.

D.M.A, C.A.R, and M.P.S. drafted the manuscript, puzzled out the implementation of frame phases, and brainstormed effective ways to benchmark the gate.
D.M.A performed the experiments.
N.D. provided theory to describe the turn-on phase of shaped flux pulses and the formula for the coherence-limited gate fidelity. 
B.R.J., C.A.R. and M.P.S. organized the effort.

\end{acknowledgments}

\bibliography{references}

\appendix

\section{Phases of the \texorpdfstring{$\XY$}{XY} Parametric Gates}\label{app:phase_of_xy_pulse}

The $\XY$ family of parametric entangling gates is obtained from the capacitive coupling between a fixed-frequency transmon, $F$, and a tunable transmon, $T$. 
Expressed in the transmon eigenbasis (notation $\ket{F,T}=\ket{F}\otimes\ket{T}$), the Hamiltonian reads
\begin{align}
H(t) &= \omega_{F_{01}}\ket{1}\bra{1}\otimes\mathbbm{1} + \omega_{T_{01}}(t)\mathbbm{1}\otimes\ket{1}\bra{1} \nonumber\\
&+ g(\ket{1}\bra{0}+\ket{0}\bra{1})\otimes(\ket{1}\bra{0}+\ket{0}\bra{1}).
\end{align}
The tunable transmon frequency, $\omega_{T_{01}}(t)$, is modulated by applying a flux bias,
$\Phi(t)=\Phi_\mathrm{dc}+\Phi_\mathrm{ac} u(t)\cos(\omega_pt+\phi_p)$, around the DC bias $\Phi_\mathrm{dc}$ with the amplitude $\Phi_\mathrm{ac}$.
The modulation envelope, $u(t)$, is characterized by a rise time, a fall time and, in between, the interaction time $\tau$ during which $u(t)=1$.
The modulation frequency $f_p=\omega_p/2\pi$, chosen to compensate the detuning between the  transmons, depends on the modulation amplitude. 
The parametric gates are thus activated during the interaction time, $t_\mathrm{rise}<t<t_\mathrm{rise}+\tau$,
and the transmons can be considered uncoupled during the rise and fall times.
The angles $\theta$ and $\beta$ of the $\XY$ gate are defined by the parameters of the effective Hamiltonian, obtained in the interaction picture,
\begin{align}
H_\mathrm{int}(t) &= ge^{i\Delta(t)}\ket{01}\bra{10}+\mathrm{H.c.}, \\
\Delta(t) &= \int_0^t\mathrm{d}t'[\omega_{T_{01}}(t')-\omega_{F_{01}}],
\end{align}
where we have used the rotating wave approximation (the $\ket{00}\leftrightarrow\ket{11}$ transition is highly detuned relative to the coupling strength).
The transmon frequency can be Fourier expanded,
\begin{align}
\omega_{T_{01}}(t) &= \wbf_0(t) + 2\sum_{k=1}^\infty \wbf_k(t) \cos[k(\omega_pt+\phi_p)], \\
\wbf_k(t) &= \sum_{n\in\mathbb{N}} \nu_n \cos(n\phi_\mathrm{dc}+k\tfrac{\pi}{2}) \mathrm{J}_k[n\phi_\mathrm{ac} u(t)],
\end{align}
with $\phi_{\mathrm{dc},\mathrm{ac}}=2\pi\Phi_{\mathrm{dc},\mathrm{ac}}/\Phi_0$, $\mathrm{J}_k(x)$ the Bessel function and $\nu_n$ the Fourier coefficients of the tunable transmon frequency with respect to the DC flux bias~\cite{Didier:2018}.

To get a qualitative understanding of the effect of the rise time on the entangling phase, we simplify the transient dynamics by assuming that the harmonics are switched on with the profile $u(t)$.
We furthermore consider a symmetric profile ($t_\mathrm{fall}=t_\mathrm{rise}$) based on the error function, 
$u(t)=\{\erf[(t-t_1)/(\sigma t_\mathrm{rise})]-\erf[(t-t_2)/(\sigma t_\mathrm{rise})]\}/2$, with $t_1=t_\mathrm{rise}/2$, $t_2=\tau+3t_\mathrm{rise}/2$ and $\sigma=1/\sqrt{32\log2}$.
During the interaction time, $t\in[t_\mathrm{rise},t_\mathrm{rise}+\tau]$, the dynamical phase is well approximated by
\begin{align}
\Delta(t) &= (\wbf_0-\omega_{F_{01}})t + \sum_{k=1}^\infty\frac{2\wbf_k}{k\omega_p} \sin[k(\omega_pt+\phi_p)] + \alpha,\\
\alpha &= \tfrac{1}{2}(\omega_{T_{01}}-\wbf_0)t_\mathrm{rise} \nonumber\\&
- \sum_{k=1}^\infty e^{-(\tfrac{\sigma}{2} k\omega_pt_\mathrm{rise})^2} \frac{2\wbf_k}{k\omega_p} \sin[k(\tfrac{1}{2}\omega_pt_\mathrm{rise}+\phi_p)].
\label{coupling_phase}
\end{align}
The Hamiltonian $H_\mathrm{int}(t)$ can then be Fourier expanded, leading to,
\begin{align}
H_\mathrm{int}(t)=g\sum_{n\in\mathbb{Z}}\varepsilon_ne^{i(\wbf_0+n\omega_p-\omega_{F_{01}})t+i(n\phi_p+\alpha)}\ket{01}\bra{10}+\mathrm{H.c.}
\end{align}
with $\varepsilon_n\in\mathbb{R}$~\cite{Didier:2018}.
An $\XY$ gate is activated when a sideband $n_0$ is at resonance with the fixed transmon frequency, 
$\wbf_0+n_0\omega_p=\omega_{F_{01}}$.
When parking at a DC sweet spot ($\Phi_\mathrm{dc}=0$ and $\Phi_\mathrm{dc}=\Phi_0/2$ for asymmetric SQUIDs), the weight of odd sidebands vanishes, and when the fixed transmon is below the tunability band, the chosen sideband is usually $n_0=-2$.
We finally obtain the effective Hamiltonian,
\begin{align}
H_\mathrm{eff} = -g_\mathrm{eff}[e^{i\beta}\ket{01}\bra{10} + e^{-i\beta}\ket{10}\bra{01}],
\end{align}
with $g_\mathrm{eff}=g|\varepsilon_{n_0}|$ the coupling strength renormalized by the sideband weight, and $\beta=-2\phi_p+\alpha+\pi$ ($\beta=-2\phi_p+\alpha$) the phase if $\varepsilon_{n_0}>0$ ($\varepsilon_{n_0}<0$).
The corresponding evolution operator, $e^{-iH_\mathrm{eff}t}$, is equal to the $\XY$ unitary defined in \cref{unitary}, up to a sign convention for $\theta$, with $\theta=g_\mathrm{eff}(t-t_\mathrm{rise})$.

Because $\varepsilon_n$ is independent of the modulation phase, the angle $\theta$ is $\phi_p$-independent.
The dependence of the angle $\alpha$ on $\phi_p$ in ~\cref{coupling_phase} is suppressed by $e^{-(\omega_pt_\mathrm{rise}\sigma)^2}$, which is below $10^{-3}$ for $t_\mathrm{rise}>2/f_p$.
For a a typical modulation frequency of $f_p=200\,\mathrm{MHz}$, this term is negligible for rise times above $10\,\mathrm{ns}$.
Up to a constant offset, the dependence of the angle $\beta$ on the modulation phase is then well approximated by 
$\beta\equiv\mp2\phi_p$, 
depending on the sign of $\wbf_0-\omega_{F_{01}}$.

\section{iRB}\label{app:irb}

\begin{figure}
    \includegraphics[width=\linewidth]{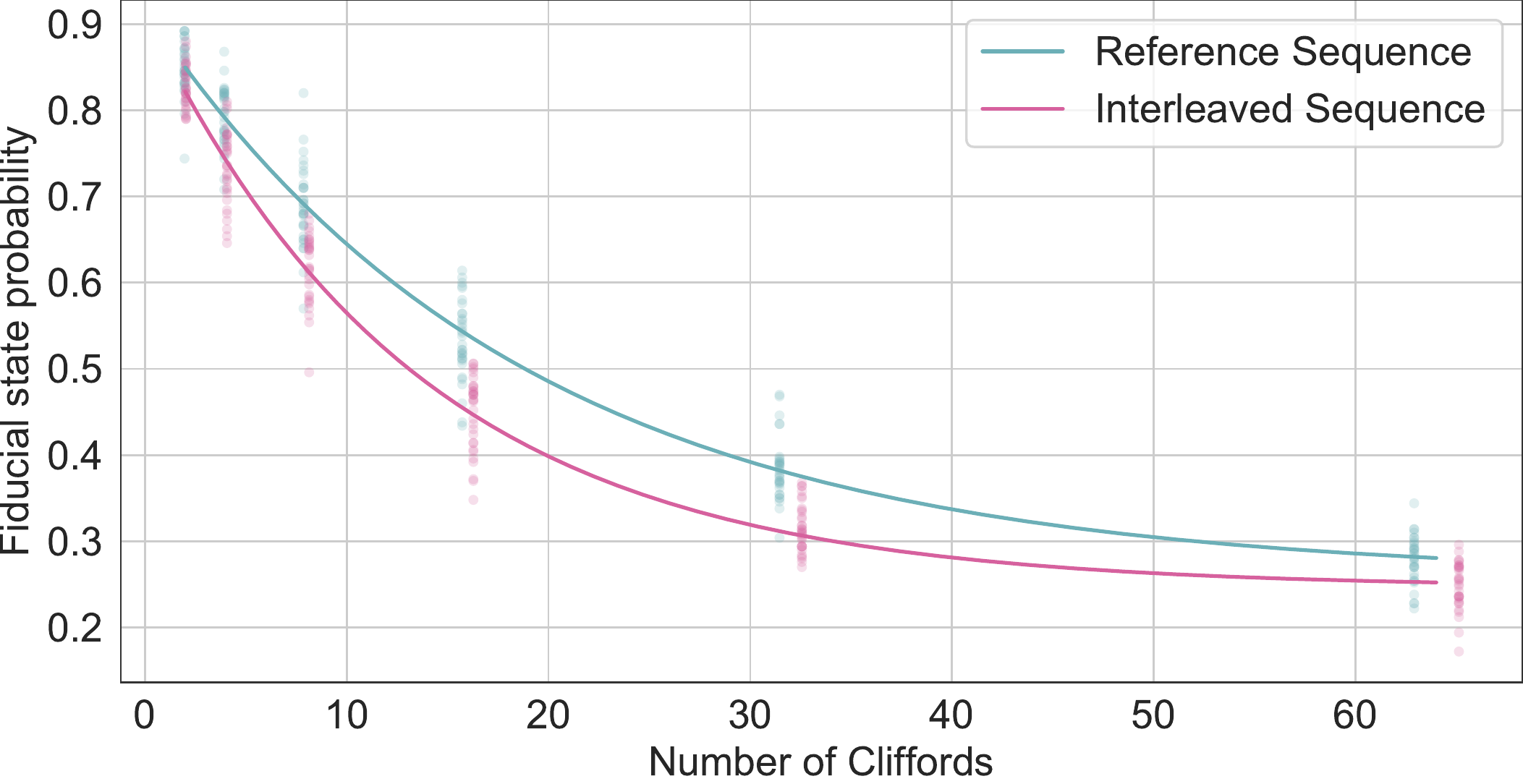}
    \caption{Representative example of iRB, performed on $\iSWAP$ constructed from a single flux pulse. Data points are the results of each individual trial. Curves are fits to an exponential decay.}
    \label{fig:irb_example}
\end{figure}

In this paper, we use interleaved randomized benchmarking (iRB) to measure the average gate fidelity of a given circuit~\cite{Magesan2012}. We construct random gate sequences with 2, 4, 8, 16, 32, and 64 Cliffords. At each sequence length, we generate 32 random circuits for the reference curve, and another 32 random circuits with an interleaved operation for the interleaved curve, for a total of $32\times6\times2=384$ different circuits. Each of these circuits is applied and measured 500 times, and results for that circuit are averaged over 500 shots. For a given iRB experiment, we apply these 384 circuits in a random order to prevent temporal fluctuations from introducing bias when comparing the reference and interleaved curves. \Cref{fig:irb_example} shows the results of one of these iRB experiments. As the circuit depth $L$ grows, the probability of measuring the expected two-qubit state decays exponentially as $B + A~p^L$ for some baseline $B$, amplitude $A$, and decay rate $p$. The average fidelity of the interleaved circuit $1-r$ can be inferred from the ratio of the decay constants for the reference curve $p$ and interleaved curve $p_{il}$ as $r = \frac{d-1}{d}(1-\frac{p_{il}}{p})$, where $d=2^n$, and $n$ is the number of qubits (which in our case is 2).

As we scale the number of $\XY$ gates used to compose an $\iSWAP$ in our iRB experiments, we expect the ratio of the decay rates to scale proportionally (i.e., we assume the errors are incoherent). Specifically, given an $\iSWAP$ composed of $n$ $\XY$ gates compared to an $\iSWAP$ composed of $m$ $\XY$ gates, we expect the ratio of the decay rates to scale as $\left(\frac{p_{il,n}}{p_n}\right)^\frac{1}{n} = \left(\frac{p_{il,m}}{p_m}\right)^\frac{1}{m}$, if the errors were due to a depolarizing channel. Both the reference and interleaved curves have an $m,n$ dependence because the scaled number of gates was used for both. Coherent errors will, in general, not follow this simple relationship.

We scale the decay ratio rather than the fidelity itself in order to handle the dimensional factors. For example, we expect that a large number of single-qubit gates will decay to a 50\% fidelity, while a large number of two-qubit gates will decay to a 25\% fidelity, which scaling the decay ratio preserves. 

\end{document}